%% file: main.tex
\documentclass[letterpaper]{article}
\usepackage{aaai2027}
\usepackage[hyphens]{url}
\usepackage{graphicx}
\usepackage{natbib}
\usepackage{caption}
\usepackage{algorithm}
\usepackage{algorithmic}

\DeclareCaptionStyle{ruled}{labelfont=normalfont,labelsep=colon,strut=off}

\usepackage{booktabs}

\copyrighttext{Preprint.}
\title{AgentRadio: Passive Awareness for Long-Horizon Multi-Agent Collaboration}
\author{
    Xinxing Ren\textsuperscript{\rm 1}\equalcontrib,
    Qianbo Zang\textsuperscript{\rm 2}\equalcontrib,
    Ziyan Wang\textsuperscript{\rm 3}\equalcontrib,
    Caelum Forder\textsuperscript{\rm 1}\equalcontrib,\\
    Suman Deb\textsuperscript{\rm 1},
    Peter Carroll\textsuperscript{\rm 1},
    Zekun Guo\textsuperscript{\rm 4}\corresponding
}
\affiliations{
    \textsuperscript{\rm 1}Coral AI Labs\\
    \textsuperscript{\rm 2}SnT, Universit\'e du Luxembourg\\
    \textsuperscript{\rm 3}King's College London\\
    \textsuperscript{\rm 4}University of Hull
}

\makeatletter
\let\agentradio@maketitle\@maketitle
\def\@maketitle{%
  {\centering
  \makebox[\textwidth][s]{%
  \includegraphics[height=0.52in]{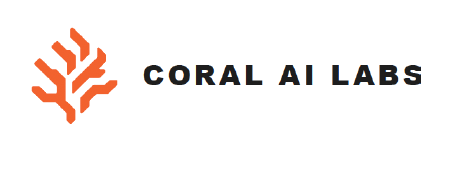}\hfill
  \includegraphics[height=0.52in]{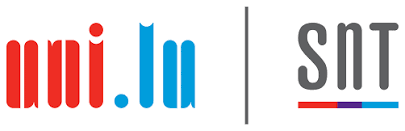}\hfill
  \includegraphics[height=0.52in]{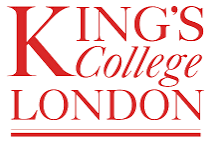}\hfill
  \includegraphics[height=0.52in]{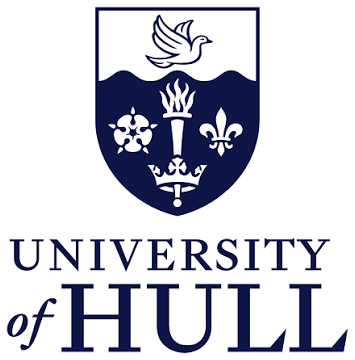}}\par}
  \vspace{12pt}
  \agentradio@maketitle}
\makeatother

\begin{document}

\maketitle

\begin{figure}[t]
\centering
\includegraphics[width=\columnwidth]{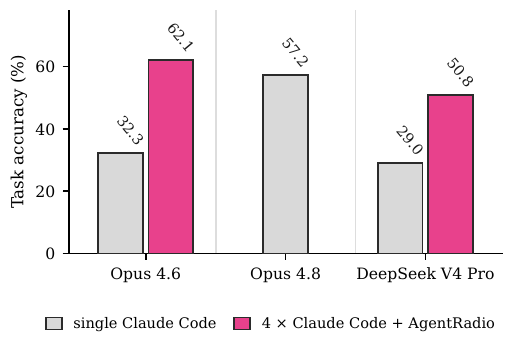}
\caption{Task accuracy on SWE-Atlas QnA (124 tasks). Four Claude Code agents organized by AgentRadio surpass the single agent of both models and the strongest single-agent entry on the public leaderboard, Claude Code with the newer Opus 4.8.}
\label{fig:fourway}
\end{figure}

\begin{abstract}
Understanding large codebases is a long-horizon task for Large Language Model (LLM) agents: answering a single question can require building and running the software, tracing execution across files, and synthesizing evidence over tens of minutes. On SWE-Atlas QnA, a benchmark of long-horizon questions over production repositories, a single Claude Code agent (Opus 4.6) resolves only 32.3\% of tasks. Dividing the work among agents with clean contexts mitigates this limitation. However, the subtasks of code comprehension are interdependent. One agent's findings can rewrite another's task, so agents must coordinate during execution, not only at phase boundaries. Existing multi-agent systems support such exchange only between phases, through staged handoffs or synchronized rounds. Communication and work remain mutually exclusive. A discovery made mid-execution cannot be shared until the next boundary. We present AgentRadio, an asynchronous message-passing layer that equips coding-agent harnesses with three primitives: threads, messages, and waiting for mentions. The last runs as a background task, surfacing teammates' messages without interrupting foreground work, so each agent remains passively aware of its peers and folds new findings into its ongoing task. Under a five-phase protocol of division of labor and negotiation, four agents organized by AgentRadio resolve 62.1\% of tasks, 29.8 points above a single agent and above Claude Code with the newer Opus 4.8 (57.2\%). Rubric-level analysis shows the gain growing with task difficulty, consistent with mid-course correction as the underlying mechanism. Our code is available at \url{https://github.com/Coral-Protocol/AgentRadio}.
\end{abstract}

\input{Sections/01_introduction}

\input{Sections/02_related_work}

\input{Sections/03_agent_radio}

\input{Sections/04_experiments}

\input{Sections/05_analysis}

\input{Sections/06_conclusion}

\bibliography{aaai2027}

\end{document}

%% file: Sections/01_introduction.tex
\section{Introduction}

The tasks handed to LLM agents keep getting longer. Benchmarks have moved from single-step question answering to navigating live websites \citep{zhou2024webarena}, operating full computer environments \citep{xie2024osworld}, resolving real GitHub issues \citep{jimenez2024swebench}, carrying out day-long professional work \citep{xu2024agentcompany}, and replicating entire research papers \citep{starace2025paperbench}. By one estimate, the length of software tasks that frontier agents can complete has doubled roughly every seven months since 2019 \citep{kwa2025measuring}. Codebase understanding is an extreme point of this arc. Answering one question about how a production system behaves requires building and running the software, tracing execution across files, and synthesizing the evidence. Yet an agent's effective attention degrades as its context grows, whether the relevant information changes position \citep{liu2024lost}, the input merely lengthens \citep{levy2024same,hsieh2024ruler}, or the interaction stretches over many turns \citep{laban2025lost}. On SWE-Atlas QnA, a benchmark of long-horizon questions over production repositories where top agents execute hundreds of commands per task \citep{raghavendra2026sweatlas}, a single Claude Code agent (Opus 4.6) resolves only 32.3\% of tasks in our runs (33.3\% on the public leaderboard), and the strongest leaderboard entry, Claude Code with Opus 4.8, reaches only 57.2\%.

\begin{figure*}[t]
\centering
\includegraphics[width=0.98\textwidth]{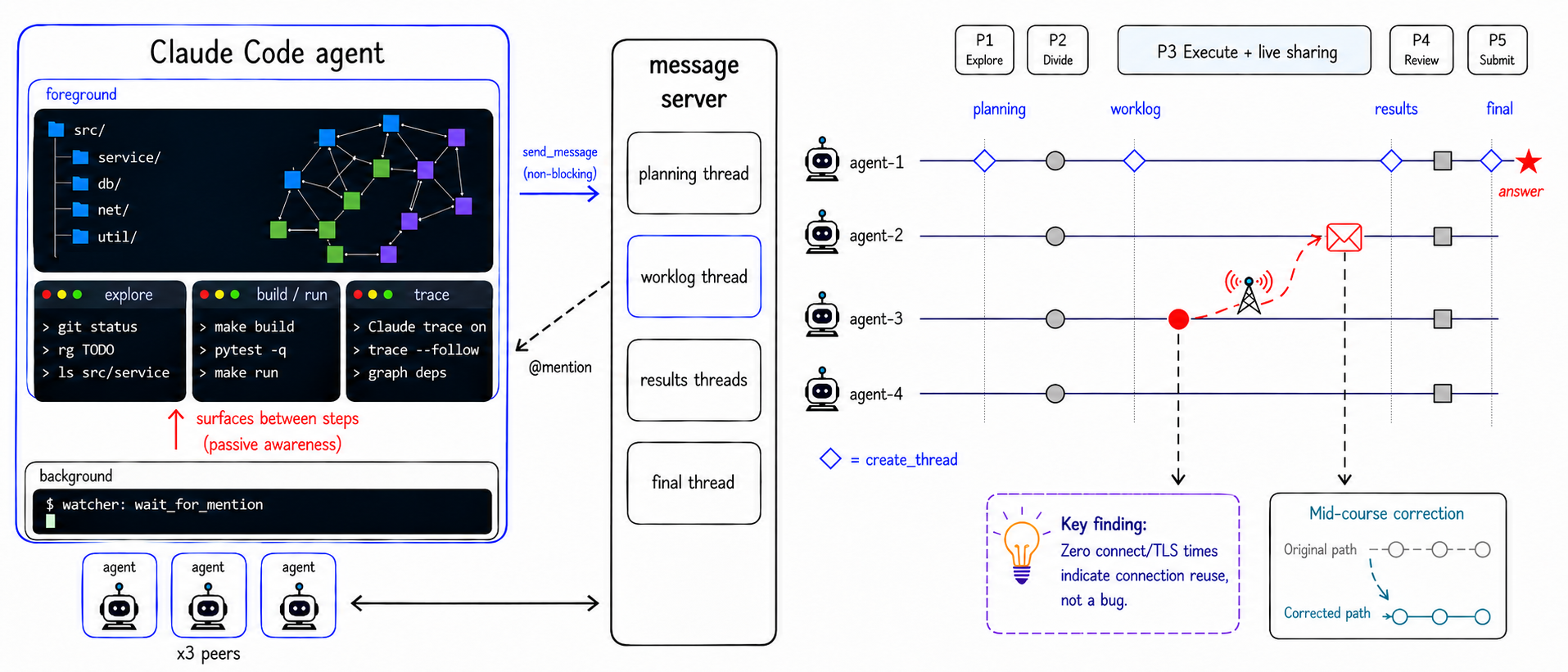}
\caption{AgentRadio and the five-phase protocol. Left: each Claude Code agent runs its harness work in the foreground while exactly one background watcher waits for mentions, so incoming messages surface between work steps and never interrupt a running command. Sending is non-blocking, and all traffic flows through named threads on a shared message server. Right: the five phases with four agent lanes. During Phase 3 an agent posts a discovery at the moment it is made, the teammate's lane never breaks, and the message is folded into the ongoing subtask at the next step. The message shown is illustrative.}
\label{fig:architecture}
\end{figure*}

When a single context cannot hold an entire task, the natural remedy is to spread the work across several agents, each starting with a clean context. This remedy pays off when the task decomposes: across 260 agent-system configurations, multi-agent architectures gained up to 80.8\% over a single agent on decomposable tasks \citep{kim2025scaling}. Anthropic reports the same pattern in production: an orchestrator delegating to parallel subagents outperformed a single agent by 90.2\% on an internal research evaluation, with much of the gain traceable to the extra tokens the parallel agents could spend \citep{anthropic2025multiagent}. When subtasks depend on one another, exchanging and reviewing intermediate conclusions helps as well. Agents that debate their answers over multiple rounds improve factuality and reasoning \citep{du2024debate,liang2024divergent}. Sampling many answers and aggregating them lifts accuracy as the number of agents grows \citep{li2024moreagents}. Learned communication topologies prune uninformative exchanges and keep the useful ones \citep{zhuge2024gptswarm,zhang2025agentprune,yu2026mpas}.

For a long task like codebase understanding, parallel division is intuitively appealing: each agent focuses on a narrower question in a cleaner context. Yet the subtasks are not independent. A fact one agent uncovers can redirect what a teammate should be doing. Therefore, negotiation has to happen in real time while the agents are working. Such real-time coordination is rare. A survey of more than 1,400 papers on LLM multi-agent systems finds asynchronicity "typically absent", with only 22 addressing it \citep{lamalfa2025missing}. The systems that do engage parallelism or asynchrony fall into three patterns. (a) Parallel but isolated: Anthropic's research system runs subagents in parallel but waits for each batch to complete and cannot steer them mid-task \citep{anthropic2025multiagent}; SoA's agents generate their components without exchanging a word \citep{ishibashi2024soa}; and CAID isolates workers in separate worktrees, coordinating only at delegation and merge \citep{geng2026caid}. (b) Parallel but round-synchronized: MPAS parallelizes message passing inside each round yet synchronizes between rounds \citep{yu2026mpas}, and debating agents exchange answers only at round boundaries \citep{du2024debate}. (c) Asynchrony in adjacent forms only: DynTaskMAS schedules sub-task execution asynchronously without inter-agent communication \citep{yu2025dyntaskmas}, HyperAgent's message queues dispatch tasks from the planner to workers with no lateral channel \citep{phan2024hyperagent}, HANDRAISER interrupts messages by interrupting the recipient delivery on communication-centric tasks \citep{wang2026handraiser}, CodeCRDT coordinates through the observation of shared state with no messages at all \citep{pugachev2025codecrdt}, and the agents of \citet{qu2026coral} append notes to a shared memory that a teammate sees only when it actively reads the store. Across all of these, an agent that is working cannot also be listening. To our knowledge, no existing system gives concurrently working agents passive awareness of one another over a lateral, natural-language channel.

We close this gap with \textbf{AgentRadio}, an asynchronous message-passing layer that equips coding-agent harnesses with three primitives: threads, messages, and waiting for mentions. The third primitive is the critical one: an agent waits for mentions as a background task, so teammates' messages surface between its work steps without interrupting the foreground, much like a radio reaches a driver whose hands never leave the wheel. Sending is likewise non-blocking. Together, these primitives keep each agent passive aware of its peers and fold incoming findings into its ongoing work. Under a five-phase protocol of division of labor and negotiation, four Claude Code (Opus 4.6) agents organized by AgentRadio resolve 62.1\% of tasks on SWE-Atlas QnA (Figure~\ref{fig:fourway}), 29.8 points above a single agent and above Claude Code with the newer Opus 4.8 (57.2\%).

\noindent Our contributions are as follows:
\begin{itemize}
\item \textbf{AgentRadio.} An asynchronous message-passing layer for coding-agent harnesses, built from three primitives (\texttt{create\_thread}, \texttt{send\_message}, \texttt{wait\_for\_mention}), whose backgrounded third member turns non-blocking messaging into passive awareness, removing the mutual exclusion between communication and work. With AgentRadio, four Claude Code agents outperform one working alone by 29.8 points, a 92\% relative gain.
\item \textbf{A controlled two-model ablation.} On 124 tasks and 1{,}306 rubrics, every layer helps on both models, and the passive step alone adds 10.5 points with Opus 4.6 and 11.3 with DeepSeek V4 Pro. Its increment is measured by changing only the communication primitives and one phase prompt.
\item \textbf{Analysis.} Rubric-level attribution of every gain and loss, a difficulty stratification showing the benefit of passive awareness growing with task difficulty, and a case study of a real mid-execution correction.
\item \textbf{Evidence of structure over compute.} The full stack beats compute-matched best-of-6 sampling (37.9\% and 31.4\% on the Opus 4.6 and DeepSeek V4 Pro).
\end{itemize}

%% file: Sections/02_related_work.tex
\section{Related Work}

\subsection{SWE-Atlas QnA}

\begin{figure}[t]
\centering
\includegraphics[width=\columnwidth]{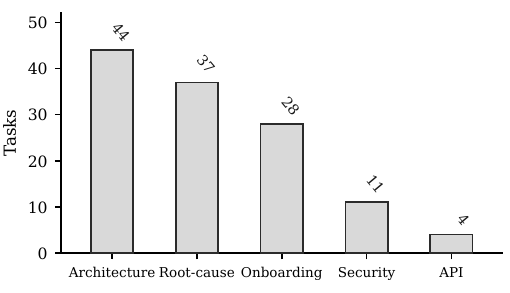}
\caption{The task mix of SWE-Atlas QnA: 124 questions over 11 production repositories in four languages \citep{raghavendra2026sweatlas}.}
\label{fig:categories}
\end{figure}

SWE-Atlas evaluates coding agents across complementary software-engineering skills, and its first released leaderboard, Codebase QnA, targets the upstream one: understanding a system deeply before changing any of its code \citep{raghavendra2026sweatlas}. The benchmark asks 124 expert-written questions about 11 production codebases in four languages. The repositories come from SWE-Bench Pro \citep{deng2025swebenchpro} and resist training-data contamination through copyleft licenses and commit-pinned Docker images. Questions are deliberately underspecified natural language, and static exploration is insufficient by design: an agent must build the software, run it with real data, and trace execution across files to answer. Figure~\ref{fig:categories} shows the task mix. The difficulty is substantial. Frontier models that exceed 80\% on SWE-bench \citep{jimenez2024swebench} resolve roughly 35\% of these tasks. Grading is equally strict. Each task carries 12.3 rubrics on average, every rubric an atomic, self-contained factual statement retained only after a three-expert majority vote on top of two rounds of human review. An LLM judge scores each rubric as a binary pass, a task counts as resolved only when every rubric passes, and modifying source files fails the task automatically.

\subsection{Coding Agents}
The tasks given to coding agents have widened alongside their tooling. Early benchmarks asked a model to complete a single function from its docstring \citep{chen2021codex}. SWE-bench moved the unit of work into a real repository, turning a GitHub issue into a patch \citep{jimenez2024swebench}. The frontier now asks for arbitrary long-horizon work inside living codebases, from multi-file changes that take a professional engineer hours to days \citep{deng2025swebenchpro} to the build, run, and comprehension tasks of SWE-Atlas \citep{raghavendra2026sweatlas}. The agents evolved in three generations. The first arranged several LLM roles into a fixed pipeline: MetaGPT and ChatDev pass structured documents from product manager to architect to engineer, so the intelligence lives in the workflow rather than in any agent's autonomy, and the environment is touched only when generated code is finally executed \citep{hong2024metagpt,qian2024chatdev}. SWE-agent replaced the pipeline with a single agent in a reason-and-act loop \citep{yao2023react} and moved the design effort into the agent-computer interface: a windowed file viewer, lint-guarded edits, and repository search that let the model act directly on what it observes \citep{yang2024sweagent}. LocAgent and RepoMaster extend the same interface-first philosophy to repository structure, turning the codebase into graphs an agent can navigate \citep{chen2025locagent,wang2025repomaster}. Today's harnesses, Claude Code, Codex CLI, and OpenCode, complete the trajectory with full shell access, persistent project memory, extensible toolsets, and models tuned to their own scaffolds \citep{claudecode2025,codexcli2025,opencode2025,raghavendra2026sweatlas}. AgentRadio takes this third generation as its unit of composition. We design neither a new workflow nor a new interface. We connect finished harnesses and change only when they can hear one another.

\subsection{When Do Multiple Agents Help?}
Early results suggested the answer was always. MetaGPT turned a one-line requirement into working software, set the state of the art on HumanEval and MBPP, and improved with every role added \citep{hong2024metagpt}. OWL's planner-and-workers design leads open-source systems on GAIA and exceeds OpenAI's Deep Research \citep{hu2025owl}. A systematic audit then reversed the mood. Across 1{,}600 annotated traces from seven popular frameworks, Cemri et al. cataloged 14 recurring failure modes and found that the gains of multi-agent systems on popular benchmarks are often minimal \citep{cemri2025whyfail}. Kim et al. quantified when each picture applies across 260 configurations, six benchmarks, and nine models from three providers \citep{kim2025scaling}. When the task decomposes, coordination pays: on real-world financial research tasks \citep{bigeard2025financeagent}, a centralized orchestrator gains 80.8\% over the single agent. When the task is sequential, forced decomposition backfires: on Minecraft crafting plans \citep{dagan2024plancraft}, every multi-agent variant loses between 39.1\% and 70.0\%. Once a single agent exceeds roughly 45\% accuracy, adding agents may yield negative returns. Codebase understanding on SWE-Atlas QnA sits squarely in the favorable regime. The task decomposes into parallel explorations, the subtasks depend on one another, and the single-agent baseline is 32.3\%, far below saturation. The experiments that follow test this prediction.

%% file: Sections/03_agent_radio.tex
\section{AgentRadio}

\subsection{Communication Primitives}
AgentRadio exposes three operations to every agent. \texttt{create\_thread(name, participants)} opens a named conversation on the message server and returns its identifier. \texttt{send\_message(thread, content, mentions)} appends a message to a thread and returns immediately, whether or not anyone is listening. A message may @-mention specific agents. \texttt{wait\_for\_mention(timeout)} blocks until a message mentioning the caller arrives, then returns that message together with a full snapshot of every thread, so the caller never needs a second read to reconstruct context.

The layer takes no position on when an agent listens. That decision, where \texttt{wait\_for\_mention} runs, is the single degree of freedom separating the two communication modes we compare. Run in the foreground, it is a blocking receive: the agent stops working in order to listen, which is our blocking baseline. Run as a background task of the harness, it becomes passive awareness: the agent keeps working, and any mention surfaces at the next step boundary (Figure~\ref{fig:architecture}, left). Everything else, the primitives, the threads, the protocol, stays fixed. This single-bit difference is what our experiments isolate.

The distinction can be stated precisely. Write an agent's run as a sequence of steps $s_1, s_2, \ldots$, one tool call each, and let $M(t)$ be the set of teammates' messages visible to the agent when it takes step $s_t$. Blocking receive makes listening a step of its own: $M$ grows only when a step is spent on \texttt{wait\_for\_mention}, so every message heard costs a step of work. Passive awareness decouples the two: $M(t)$ contains every message sent before $s_t$, and no step is spent listening.

\subsection{The Five-Phase Protocol}
We evaluate AgentRadio under a fixed protocol of division of labor and negotiation (Figure~\ref{fig:architecture} right, Algorithm~\ref{alg:protocol}). One agent, agent-1, additionally serves as the assembler. It opens the planning, worklog, and final-answer threads and gates every transition: a phase ends only after agent-1 collects an explicit approval from every agent.

\begin{algorithm}[!htb]
\caption{The five-phase protocol}
\label{alg:protocol}
\begin{algorithmic}[1]
\REQUIRE task $T$, repository $R$, agents $a_1, \dots, a_4$ with assembler $a_1$
\ENSURE final answer $F$
\STATE every agent starts its background watcher \COMMENT{P1, explore}
\STATE every agent explores $R$ and drafts the sub-questions of $T$
\STATE $a_1$ opens the planning thread \COMMENT{P2, divide}
\REPEAT
    \STATE agents share findings, propose or amend the partition
\UNTIL{$a_1$ counts an approval from every agent}
\STATE $a_1$ broadcasts the plan and opens the worklog thread
\FOR{\textbf{each} agent \textbf{in parallel}}
    \STATE work own sub-questions \COMMENT{P3, execute}
    \STATE \textbf{on} discovery that bears on a teammate, contradicts the plan, or blocks an approach: post it to the worklog
    \STATE \textbf{on} incoming message: fold it in at the next step
\ENDFOR
\STATE every agent opens a results thread and posts its findings with evidence \COMMENT{P4, review}
\REPEAT
    \STATE reviewers post conflicts, thin evidence, and unmentioned observations, reopening P3 work if needed
\UNTIL{$a_1$ counts an approval from every agent in every results thread}
\STATE $a_1$ posts a draft of $F$ in the final-answer thread \COMMENT{P5, submit}
\REPEAT
    \STATE agents post approvals or change requests and $a_1$ revises the draft
\UNTIL{$a_1$ counts four approvals}
\STATE $a_1$ submits $F$
\end{algorithmic}
\end{algorithm}
\begin{itemize}
\item \textbf{P1, explore.} Every agent starts its background watcher, independently explores the repository, and drafts the sub-questions it sees. Nothing is sent.
\item \textbf{P2, divide.} The assembler opens a planning thread. The agents pool their Phase-1 findings, negotiate a partition of the sub-questions, and revise it until every agent approves. Three events trigger a message: proposing or amending the partition, reporting a finding that conflicts with a teammate's proposal, and casting an approval.
\item \textbf{P3, execute.} Each agent works its own sub-questions. Under passive awareness, a discovery triggers a worklog post at the moment it is made: a finding that bears on a teammate's sub-question, a contradiction with the agreed plan, an obstacle, or an abandoned dead end.
\item \textbf{P4, review.} Each agent broadcasts its findings with evidence in its own results thread. A review message is triggered by a factual conflict with the reviewer's own evidence, by evidence too thin to support a claim, or by something the reviewer saw in its own work that the findings do not mention. Teammates can send a sub-question back to P3.
\item \textbf{P5, submit.} The assembler composes the final answer from the approved results, broadcasts the draft for a last round of approvals, and submits.
\end{itemize}
Under blocking receive, the same five phases run unchanged, but the live sharing of P3 disappears: hearing a message costs a foreground wait, so agents fall silent while they work, and a discovery cannot reach a teammate before P4.

\begin{table*}[t]
\centering
\small
{\footnotesize B0 = single Claude Code \quad L1 = $4\times$ Claude Code + division of labor \quad L2 = L1 + negotiation \quad L3 = L2 + passive awareness (AgentRadio)\par}
\vspace{3pt}
\begin{tabular}{@{}lcccccccc@{}}
\toprule
 & \multicolumn{4}{c}{Opus 4.6} & \multicolumn{4}{c}{DeepSeek V4 Pro} \\
\cmidrule(lr){2-5}\cmidrule(l){6-9}
 & B0 & L1 & L2 & L3 & B0 & L1 & L2 & L3 \\
\midrule
Architecture and system design (44) & 15 & 13 & 24 & \textbf{30} & 14 & 13 & 17 & \textbf{24} \\
Root-cause analysis (37)            & 9  & 16 & 18 & \textbf{20} & 11 & 13 & 15 & \textbf{18} \\
Code onboarding (28)                & 11 & 12 & 14 & \textbf{18} & 7  & 8  & 10 & \textbf{13} \\
Security (11)                       & 4  & \textbf{7} & \textbf{7} & \textbf{7} & 4 & 4 & 6 & \textbf{7} \\
API and library integration (4)     & 1  & 1  & 1  & \textbf{2}  & 0  & 0  & 1  & \textbf{1} \\
\midrule
All tasks resolved (124)            & 40 & 49 & 64 & \textbf{77} & 36 & 39 & 49 & \textbf{63} \\
Task accuracy (\%)                  & 32.3 & 39.5 & 51.6 & \textbf{62.1} & 29.0 & 31.4 & 39.5 & \textbf{50.8} \\
Rubric pass rate (\%)               & 84.2 & 86.1 & 91.3 & \textbf{93.1} & 81.2 & 83.7 & 85.9 & \textbf{90.2} \\
\midrule
L3 vs.\ L2 (exact McNemar) & \multicolumn{4}{c}{wins 15, loses 2, $p=0.0023$} & \multicolumn{4}{c}{wins 17, loses 3, $p=0.0026$} \\
\bottomrule
\end{tabular}
\caption{Main results on SWE-Atlas QnA (124 tasks, 1{,}306 rubrics). Category rows give tasks resolved, with category sizes in parentheses. Within each model column every configuration uses the same harness and settings. The McNemar row tests the passive-awareness step on paired task outcomes.}
\label{tab:main}
\end{table*}

\subsection{Implementation}
AgentRadio ships as two parts. The message server of Figure~\ref{fig:architecture} is a standalone process that stores threads, messages, and mentions for a group of agents and implements the three primitives. Each agent reaches it through three thin shell scripts, one per primitive. On the harness side the requirements stay minimal: the harness must only be able to run a shell command as a background task, which mainstream coding harnesses already provide, and the harness itself is never modified. The agent is instructed in its prompt to keep one watcher running and to send through the scripts. Switching from blocking to passive receive therefore adds no LLM calls, because the watcher is an ordinary operating-system process rather than an agent step: the only new tokens an agent pays for are the messages that surface.

%% file: Sections/04_experiments.tex
\section{Experiments}
\subsection{Setup and Configurations}
All experiments run on the 124 tasks of SWE-Atlas QnA under the benchmark's rules: each agent works in the sandboxed container against the same repository checkout, may build and run the software, and may not modify source files. We report two metrics. Task accuracy counts a task as resolved only when every one of its rubrics passes. Rubric pass rate counts the share of all 1{,}306 rubrics passed and moves in finer steps. Grading uses the benchmark's own LLM judge (Claude Opus 4.5), fixed across every configuration.

We compare five configurations. \textbf{B0} is a single Claude Code agent. \textbf{B1} repeats \textbf{B0} six times independently, spending six single-agent budgets, and reports the best of the six complete runs. \textbf{L1} moves to a team of four Claude Code agents and adds division of labor: one of the four briefly explores the repository and partitions the question, each agent then solves its sub-questions independently, and the answers are merged without review. \textbf{L2} adds negotiation: every agent first explores independently and publishes its initial findings, one agent then proposes the partition, the team reviews it until everyone approves, and the results are cross-reviewed at the end, which is the five-phase protocol of the previous section under blocking receive. \textbf{L3} runs the same protocol under passive awareness, which is the full AgentRadio configuration. The step from L2 to L3 changes only the communication mode.

Every configuration runs once per task with identical model settings (thinking effort high, temperature 0). All comparisons are therefore paired: same tasks, same harness, same model, one coordination change at a time. We repeat the whole ladder with an open-source model, DeepSeek V4 Pro, inside the same harness.

\subsection{Main Results}

Table~\ref{tab:main} reports the ladder. Every layer helps, on both models. Division alone lifts the single agent by 7.2 points with Opus 4.6 and 2.4 with DeepSeek. Negotiation adds another 12.1 and 8.1. Passive awareness adds a further 10.5 and 11.3. The passive increment is the step our design isolates, and it is statistically solid on the paired task outcomes of the McNemar row. The rubric pass rate moves the same way.

The category rows of Table~\ref{tab:main} show where each layer earns its points. Architecture and system design, the questions that span the most components, is the one category where naive division loses tasks on both models, 15 to 13 with Opus 4.6 and 14 to 13 with DeepSeek. Negotiation recovers the loss and passive awareness extends it, which makes architecture the largest gainer of the full stack, up 15 tasks with Opus 4.6 and 10 with DeepSeek. Root-cause analysis behaves in the opposite way: division alone lifts Opus 4.6 from 9 to 16 resolved tasks, consistent with candidate hypotheses that agents can explore independently. The passive step itself pays most where questions cut across the partition, in architecture (+6 and +7 tasks) and onboarding (+4 and +3).

\subsection{Compute and Model-Generation Baselines}

\begin{table}[t]
\centering
\footnotesize
\setlength{\tabcolsep}{4pt}
\begin{tabular}{@{}lcccc@{}}
\toprule
 & \multicolumn{2}{c}{Opus 4.6} & \multicolumn{2}{c}{DeepSeek V4 Pro} \\
\cmidrule(lr){2-3}\cmidrule(l){4-5}
Configuration & Acc. & \$/task & Acc. & \$/task \\
\midrule
B0 \enspace single agent             & 32.3 & 2.96  & 29.0 & 0.42 \\
B1 \enspace best of six single runs  & 37.9 & 17.76 & 31.4 & 2.52 \\
L1 \enspace four agents + division   & 39.5 & 5.38  & 31.4 & 0.77 \\
L2 \enspace + negotiation            & 51.6 & 15.59 & 39.5 & 1.93 \\
L3 \enspace AgentRadio (full stack)  & \textbf{62.1} & 19.45 & \textbf{50.8} & 2.46 \\
\midrule
Single Opus 4.8 & 57.2 & -- & -- & -- \\
\bottomrule
\end{tabular}
\caption{Task accuracy (\%) and average API cost per task (USD) for every configuration. B1's cost is the price of its six independent single-agent runs. Single Opus 4.8 is the strongest single-agent entry on the public leaderboard.}
\label{tab:compute}
\end{table}

Coordination is not free, and Table~\ref{tab:compute} prices it: the full stack spends \$19.45 per task with Opus 4.6 and \$2.46 with DeepSeek, about six times a single agent's spend, so the gains of Table~\ref{tab:main} could in principle come from budget rather than coordination. B1 tests this. It spends six single-agent budgets (\$17.76 and \$2.52), nearly the full stack's spend, on six independent runs and keeps the best, reaching 37.9\% with Opus 4.6 and 31.4\% with DeepSeek. Naive division lands at the same level for less than a third of B1's price, and neither comes close to the full stack. Budget alone buys 5.6 and 2.4 points over the single agent, and at essentially the same spend the full stack returns a further 24.2 and 19.4. Within the stack, the passive step is the economical one, a quarter more spend than blocking negotiation for 10.5 and 11.3 points. Four Opus 4.6 agents under AgentRadio also surpass a single agent running the newer Opus 4.8, 62.1\% against 57.2\%, the strongest single-agent entry on the public leaderboard.

\subsection{Run-to-Run Variance}

\begin{table}[t]
\centering
\footnotesize
\setlength{\tabcolsep}{5pt}
\begin{tabular}{@{}lccccc@{}}
\toprule
Configuration & Run 1 & Run 2 & Run 3 & Mean & SD \\
\midrule
B0             & 40.0 & 30.0 & 36.7 & 35.6 & 5.1 \\
L1   & 43.3 & 43.3 & 53.3 & 46.7 & 5.8 \\
L2            & 53.3 & 56.7 & 50.0 & 53.3 & 3.4 \\
L3  & \textbf{63.3} & \textbf{63.3} & \textbf{66.7} & \textbf{64.4} & \textbf{2.0} \\
\bottomrule
\end{tabular}
\caption{Task accuracy (\%) of three independent runs per configuration on a 30-task subset (9 architecture, 9 root-cause, 6 onboarding, 3 security, 3 API tasks, 318 rubrics, Opus 4.6).}
\label{tab:variance}
\end{table}

Table~\ref{tab:variance} bounds the run-to-run noise of the single-run protocol: on a 30-task subset of 318 rubrics, three independent runs per configuration with Opus 4.6 preserve the ordering of the ladder, and the worst of the three passive runs at 63.3\% still beats the best run of every other configuration at 56.7\%.

%% file: Sections/05_analysis.tex
\section{Analysis}
This section analyzes the Opus 4.6 runs at the rubric level.

\subsection{Gross and Net Attribution}

\begin{figure}[t]
\centering
\includegraphics[width=\columnwidth]{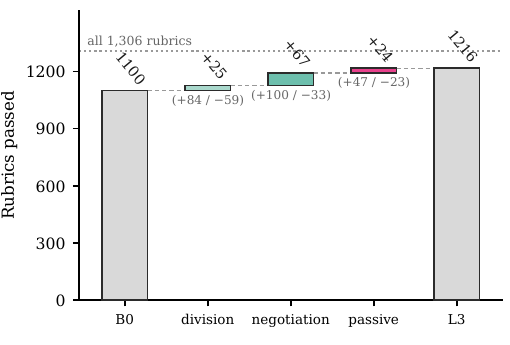}
\caption{Rubric-level attribution of the Opus 4.6 ladder, as a waterfall from the single agent to the full protocol. Each floating bar gives the net change of one layer, with the gross rubrics gained and lost below it. The dotted line marks the benchmark total of 1{,}306 rubrics.}
\label{fig:waterfall}
\end{figure}

\begin{figure}[t]
\centering
\includegraphics[width=\columnwidth]{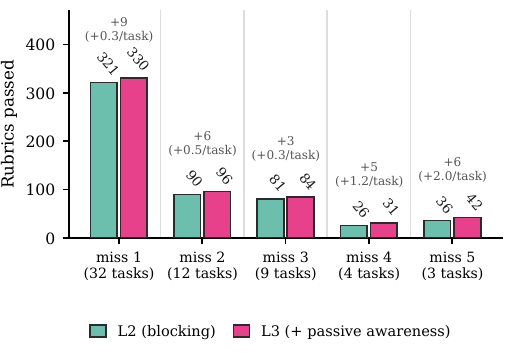}
\caption{Rubrics passed on the 60 tasks that L2 leaves unresolved (Opus 4.6), grouped by how many rubrics L2 misses. Annotations give each group's gain in total and per task.}
\label{fig:difficulty}
\end{figure}

Table~\ref{tab:main} reports what each layer adds on balance, and the balance hides churn. Figure~\ref{fig:waterfall} splits every step of the Opus 4.6 ladder into the rubrics it gains and the rubrics it loses. Division is the noisiest step. It gains 84 rubrics that the single agent misses and loses 59 that the single agent passes, a net of +25. Partitioning lets four agents cover ground that one agent cannot, and it also cuts questions that span the partition into fragments, the failure that surfaces as the architecture regression in Table~\ref{tab:main}. Negotiation is the largest and the cleanest contributor. It adds 100 gross rubrics against 33 lost, a net of +67. Joint exploration before partitioning attacks the fragmentation at its source. Passive awareness gains 47 rubrics and loses 23, a net of +24. A plausible source of the losses is the messages themselves. A message that arrives mid-execution can pull an agent off a line of evidence that would have passed a rubric, but the gains outweigh these losses, as the next subsection shows.

\subsection{Gain by Task Difficulty}
\label{sec:difficulty}

Figure~\ref{fig:difficulty} takes the 60 tasks that L2 leaves unresolved on the Opus 4.6 runs and groups them by how many rubrics L2 still misses. The gain from passive awareness grows along this axis. On the near-miss groups that miss one to three rubrics, it adds between 0.3 and 0.5 rubrics per task. On the four tasks that miss four rubrics it adds 1.2 rubrics per task, and on the three tasks that miss five it adds 2.0. The harder the task is for the blocking protocol, the more passive awareness contributes. On a near miss only a few rubrics remain, so the headroom for any correction is small, and passive awareness still recovers part of it. A wide miss means the plan itself is wrong or the decisive evidence sits across sub-question boundaries, the failures a mid-execution message can catch while there is still time to act.

\subsection{Case Studies}

\begin{figure}[t]
\centering
\includegraphics[width=\columnwidth]{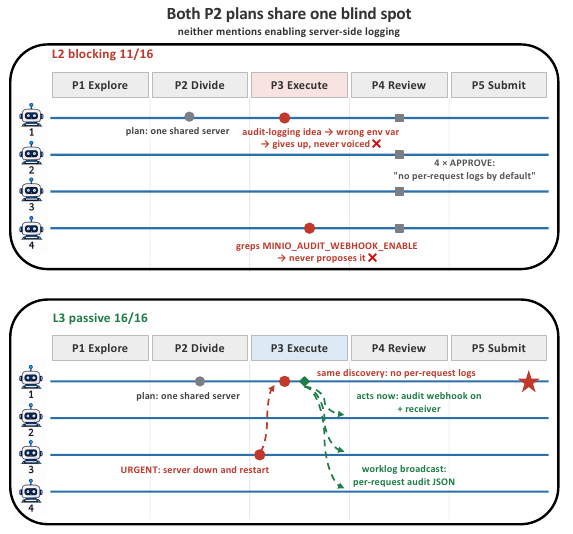}
\caption{The MinIO case with Opus 4.6. Neither Phase 2 plan mentions enabling server-side logging. Under blocking receive (A) the discovery dies unvoiced and the review approves the wrong consensus. Under passive awareness (B) the same discovery leads to instrumenting the shared server and broadcasting per-request evidence, and all five logging rubrics flip.}
\label{fig:caseminio}
\end{figure}

\begin{figure}[!tb]
\centering
\includegraphics[width=\columnwidth]{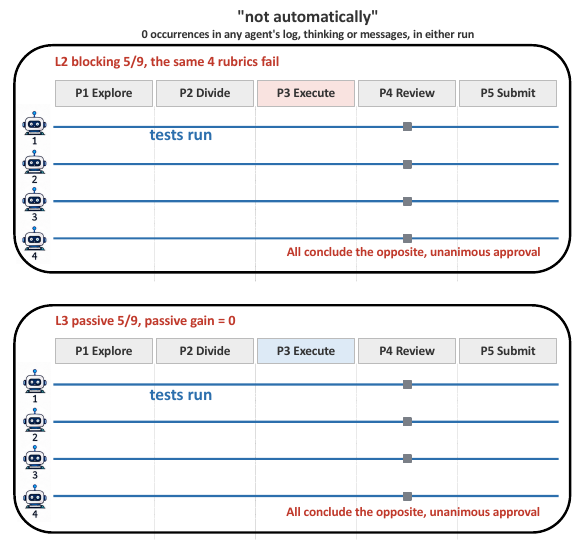}
\caption{The Grafana case with Opus 4.6. The two runs mirror each other: both teams run the tests, no agent states the negative conclusion that four rubrics require, and the same four rubrics fail, a passive gain of zero.}
\label{fig:casegrafana}
\end{figure}

In a MinIO task (Figure~\ref{fig:caseminio}), five of 16 rubrics require server-side per-request evidence that the console does not print by default. The two Phase 2 plans share the same blind spot, neither mentions enabling server-side logging, so the need can only surface during execution. In the blocking run the discovery happens twice and dies twice. Agent-1 privately considers audit logging, fails with a wrong environment variable, and gives up without a word. Agent-4 greps the exact switch, \texttt{MINIO\_AUDIT\_WEBHOOK\_ENABLE}, and never proposes it. The review then unanimously approves ``no per-request logs by default'', since Phase 3 gives a discovery no outlet and Phase 4 never asks what an agent privately ruled out. The passive run makes the same discovery and acts at once: agent-1 enables the audit webhook on the team's shared server and broadcasts the per-request records to the worklog, so one instrumentation becomes team-wide evidence and the score moves from 11 of 16 to 16 of 16.

A Grafana provisioning task (Figure~\ref{fig:casegrafana}) bounds what the layer can do. Four of nine rubrics require negative conclusions, for example that the datasource picker does not auto-select, and the phrase ``not automatically'' appears zero times in any agent's log in either run. Both teams run the relevant tests, every agent still concludes the opposite, and both configurations pass 5 of 9 rubrics. Passive awareness turns discoveries that agents already make, and would otherwise swallow, into team-wide evidence. It cannot supply a conception that no agent forms.

%% file: Sections/06_conclusion.tex
\section{Conclusion}
We presented AgentRadio, an asynchronous message-passing layer that equips coding-agent harnesses with three primitives and one new capability: run as a background task, the wait primitive keeps an agent passively aware of its teammates. Under a five-phase protocol of division of labor and negotiation, four Claude Code agents organized by AgentRadio raise task accuracy on SWE-Atlas QnA from 32.3\% to 62.1\% with Opus 4.6 and from 29.0\% to 50.8\% with DeepSeek V4 Pro. The passive step survives a paired significance test on both models and pays most where blocking misses worst. It changes neither the model, the harness, nor the protocol, only when agents can hear one another, a single degree of freedom worth more than a model generation.